\documentclass[12pt]{article}
\usepackage{amsmath, amsfonts, amssymb, amsthm, enumerate, graphicx}
\usepackage{epstopdf}
\usepackage{amsopn}
\usepackage{mathrsfs}

\newtheorem{theorem}{Theorem}[section]
\newtheorem{definition}[theorem]{Definition}

\def\bW{\mathcal{W}}
\def\FF{\mathcal{F}}

\def\DD{\mathcal{D}}
\def\TT{\mathcal{T}}
\renewcommand{\baselinestretch}{1.5}

\begin{document}
\renewcommand{\baselinestretch}{1.5}
\title{An upper bound on the dimension of the voting system of the European Union Council under the Lisbon rules}
\author{Yuming Chen,  Wai Shun Cheung and Tuen Wai Ng\footnote{Department of Mathematics, The University of Hong Kong, Pokfulam, Hong Kong. \newline \quad E-mail address:chenym@connect.hku.hk,\, cheungwaishun@gmail.com,\, ntw@maths.hku.hk}}

\maketitle

\begin{abstract}
The voting rules of the European Council (EU) under the Treaty of Lisbon became effective on 1 November 2014. Kurz \& Napel (2015) showed that the dimension of this voting system is between $7$ and $13,368$.  The lower bound $7$ actually sets a new world record for the dimension of the real-world voting bodies. In this article, by finding a new way to represent the union of two weighted games as an intersection of certain weighted games (Theorem 1), we greatly reduce the upper bound $13,368$ to just $25$. We also consider what will happen to our upper bound and Kurz \& Napel's lower bound if the United Kingdom is no longer a member of the European Union Council. 
\end{abstract}

%
 
\section{Introduction}
The voting rules used in the European Council (EU) under the Treaty of Lisbon (which became effective on 1 November 2014) is quite unique within the current, global range of electoral systems. It not only requires a ``qualified majority'' of both the number of member states supporting a proposal and the population of the supporting member states, but also specifies a ``blocking minority'' which can block a proposal if certain condition is satisfied. It is the existence of such  a ``blocking minority'' that makes the system complicated and interesting to study from the mathematical point of view. In fact, Kurz \& Napel (2015) showed that this voting system has dimension between $7$ and $13368$ and therefore this dimension sets a world record of the dimension of the real-world voting bodies. Indeed, the previous record holders are EU under Treaty of Nice and the Legislative Council of Hong Kong and both of them have dimension $3$ (see Freixas (2004) and Cheung \& Ng (2014)).  In this article, we will reduce Kurz \& Napel's upper bound to $25$ or even $24$ (depending on the populations of the countries in EU). This is achieved by finding a new way to represent the union of two weighted games as an intersection of certain weighted games (Theorem 1). It is expected that Theorem 1 will be useful for computing the dimension of other real world voting systems. Finally, assuming that the United Kingdom is no longer a member of the European Union Council, we will show that our upper bound will jump to $1362$ while Kurz \& Napel's lower bound will only increase to $8$.

\section{Notations and definitions}
\begin{definition}
A {\it simple game} is a pair $(N, v)$  where $N$ is the set of players described as $\{1, \ldots , n\}$ and $v$ is the characteristic 
function which satisfies:
\begin{enumerate}
\item $v(S)=0$ or $1$, for all $S\subseteq N$;
\item $v(S)\le v(T)$ if $S\subseteq T$;
\item $v(\emptyset)=0$ and $v(N)=1$.
\end{enumerate}
A coalition $S$ is {\it winning} if $v(S)=1$ and {\it losing} if $v(S)=0$ and we let $\bW(v)$ be the set of winning coalitions of $v$.
\end{definition}
\begin{definition} A {\it weighted majority game} is a simple game which can be realized by a vector $(w_1 , . . . , w_n )$ together with a threshold $q$ which makes
the representation $[q; w_1,\ldots,w_n]$ so that $S$ is a winning coalition if and only if $\sum_{j\in S}w_j\ge q$.
\end{definition}

\begin{definition} Let $v_1,v_2$ be two simple games with identical player set $N$.  The intersection of $v_1$ and $v_2$, denoted by $v_1\wedge v_2$ is the simple game with $\bW(v_1\wedge v_2)=\bW(v_1)\cap \bW(v_2)$ and the union of $v_1$ and $v_2$, denoted by $v_1\vee v_2$ is the simple game with $\bW(v_1\vee v_2)=\bW(v_1)\cup \bW(v_2)$.
\end{definition}

\begin{definition} The {\it dimension} of $v$ is the smallest $k$ such that $v$ coincides with the intersection $v_1\wedge v_2\cdots \wedge v_k$  of $k$ weighted games.

The {\it codimension} of $v$ is the smallest $k$ such that $v$ coincides with the union $v_1\vee v_2\cdots\vee v_k$ of $k$ weighted games.

The {\it boolean dimension} of $v$ is the smallest $k$ such that $v$ can be represented as unions and intersections of $k$ weighted games.
\end{definition}
\newpage
\begin{table}
\footnotesize
\parbox{.45\linewidth}{
\centering
\caption{2014 EU population}
\begin{tabular}{llr}
\# &Country & Population \\
1&Germany	&80,780,000 \\
2&France&	65,856,609\\
3&United Kingdom&	64,308,261\\
4&Italy &	60,782,668\\
5&Spain&	46,507,760\\
6&Poland&	38,495,659\\
7&Romania&	19,942,642\\
8&Netherlands&	16,829,289\\
9&Belgium&	11,203,992\\
10&Greece&	10,992,589\\
11&Czech Republic&	10,512,419\\
12&Portugal&	10,427,301\\
13&Hungary&	9,879,000\\
14&Sweden&	9,644,864\\
15&Austria&	8,507,786\\
16&Bulgaria&	7,245,677\\
17&Denmark&	5,627,235\\
18&Finland&	5,451,270\\
19&Slovakia&	5,415,949\\
20&Ireland&	4,604,029\\
21&Croatia&	4,246,700\\
22&Lithuania&	2,943,472\\
23&Slovenia&	2,061,085\\
24&Latvia&	2,001,468\\
25&Estonia&	1,315,819\\
26&Cyprus&	858,000\\
27&Luxembourg&	549,680\\
28&Malta&	425,384\\
&Total population&	507,416,607\\
\end{tabular}} \hfill
\parbox{.45\linewidth}{
\centering
\caption{2016 EU population}
\begin{tabular}{llr}
\#&Country & Population \\
1&Germany&	82,175,684\\
2&France&	66,730,453\\
3&United Kingdom&	65,382,556\\
4&Italy&	60,665,551\\
5&Spain&	46,440,099 \\
6&Poland&	37,967,209\\
7&Romania&	19,760,314\\
8&Netherlands&	16,979,120  \\
9&Belgium&	11,311,117 \\
10&Greece&	10,783,748\\
11&Czech Republic&	10,553,843\\
12&Portugal&	10,341,330\\
13&Sweden&	9,851,017\\
14&Hungary&	9,830,485\\
15&Austria&	8,700,471 \\
16&Bulgaria&	7,153,784\\
17&Denmark	&5,707,251\\
18&Finland&	5,487,308\\
19&Slovakia&	5,426,252\\
20&Ireland&	4,726,286 \\
21&Croatia&	4,190,669\\
22&Lithuania&	2,888,558\\
23&Slovenia&	2,064,188\\
24&Latvia&	1,968,957\\
25&Estonia&	1,315,944\\
26&Cyprus&	848,319\\
27&Luxembourg&	576,249\\
28&Malta&	450,415 \\
&Total population&	510,277,177
\end{tabular}}
\end{table}
\newpage
\begin{table}
\footnotesize
\parbox{.45\linewidth}{
\centering

\caption{2017 EU population}
\begin{tabular}{llr}
\#&Country & Population \\
1&Germany&	82,521,653 \\
2&France&	66,989,083\\
3&United Kingdom&	65,808,573 \\
4&Italy&	60,589,445 \\
5&Spain&	46,527,039  \\
6&Poland&	37,972,964 \\
7&Romania&	19,644,350 \\
8&Netherlands&	17,081,507 \\
9&Belgium&	11,351,727 \\
10&Greece&	10,768,193 \\
11&Czech Republic&	10,578,820 \\
12&Portugal&	10,309,573 \\
13&Sweden&	9,995,153 \\
14&Hungary&	9,797,561\\
15&Austria&	8,772,865 \\
16&Bulgaria&	7,101,859 \\
17&Denmark	&5,748,769\\
18&Finland&	5,503,297 \\
19&Slovakia&	5,435,343 \\
20&Ireland&	4,784,383 \\
21&Croatia&	4,154,213 \\
22&Lithuania&	2,847,904\\
23&Slovenia&	2,065,895 \\
24&Latvia&	1,950,116 \\
25&Estonia&	1,315,635 \\
26&Cyprus&	854,802 \\
27&Luxembourg&	590,667\\
28&Malta&	460,297\\
&Total population&	511,521,686 
\end{tabular}}
\parbox{.45\linewidth}{
\centering
\caption{2018 EU population}
\begin{tabular}{llr}
\#&Country & Population \\
1&Germany&	82,850,000\\
2&France&	67,221,943\\
3&United Kingdom&	66,238,007\\
4&Italy&	60,483,973 \\
5&Spain&	46,659,302\\
6&Poland&	37,976,687 \\
7&Romania&	19,523,621\\
8&Netherlands&	17,181,084 \\
9&Belgium&	11,413,058 \\
10&Greece&	10,738,868\\
11&Czech Republic&	10,610,055 \\
12&Portugal&	10,291,027 \\
13&Sweden&	10,120,242 \\
14&Hungary&	9,778,371\\
15&Austria&	8,822,267\\
16&Bulgaria&	7,050,034 \\
17&Denmark	& 5,781,190  \\
18&Finland&	5,513,130 \\
19&Slovakia&	5,443,120 \\
20&Ireland&	4,838,259\\
21&Croatia&	4,105,493 \\
22&Lithuania&	2,808,901\\
23&Slovenia&	2,066,880 \\
24&Latvia&	1,934,379 \\
25&Estonia&	1,319,133\\
26&Cyprus&	864,236\\
27&Luxembourg&	602,005 \\
28&Malta&	475,701\\
&Total population&	512,710,966
\end{tabular}}
\end{table}

\section{Data sets and the voting rule}
Let us first state the rule of the voting game $v_{EU}$ of the EU systems:
We numbered the 28 EU members by $1,\ldots,28$ (see Table $1$ or $2$).  A motion will be passed if
\begin{enumerate}
\item[(1)] at least $55\%$ of EU members support the motion (and we let the weighted majority game $v_{16}:=[16; 1,1,\ldots,1]$)

\noindent
and either  
\end{enumerate}
\begin{enumerate}
\item[(2a)] the EU members that support the motion represent at least $65\%$ of the total population (and we let $v_{65}:=[0.65\sum w_j; w_1, w_2,\ldots, w_{28}]$ where $w_j$ is the population of the $j$-th country), or
\item[(2b)] no more than four EU members reject the motion (and we let $v_{25}:=[25; 1,\ldots,1]$).
\end{enumerate}
Therefore $v_{EU}=v_{16}\wedge (v_{65}\vee v_{25})$ and hence the boolean dimension of $v_{EU}$ is $3$ (see \cite{KN}).  Note that we also have $v_{EU}=(v_{16}\wedge v_{65})\vee v_{25}$.

To compute the dimension of $v_{EU}$, we need to know the populations of the 28 EU members. Here we will use four data sets.  The 2014 data (Table 1) is from Kurz \& Napel (2015) \cite{KN}, it will provide a clear comparison between their estimation and ours. The 2016, 2017, 2018 data (Table 2,3,4) are taken from http://ec.europa.eu/eurostat on 7 March, 2019.  Indeed, according to Kurl \& Napel (2015), their data set was also taken from http://ec.europa.eu/eurostat, but it seems the website had adjusted the data afterward. Note that the orders of the countries are not the same in the four tables, as we arranged the countries in descending populations, which will make it easier to study the voting game mathematically.

\section{Realization of games as intersection of weighted games}

In this section, we explain two constructions which are useful to realize a game as intersections of weighted games.

\begin{itemize}
\item[I)] Suppose that we have two simple games $v$ and $v'$ with identical player set $N$ such that $\bW(v)\subset \bW(v')$. Let $\FF$ be the set of maximal coalitions which is winning in $v'$ but losing in $v$.  For each $S\in\FF$, define $v^S$ to be a weighted game with quota $1$ and the weight $w^S_j=1$ if $j\notin S$ and $0$ if $j\in S$.  Note that a coalition $W$ is losing in $v^S$ if and only if $W\subseteq S$. Therefore, we have
$$\bW(v)=\bW(v') \cap \bigcap_{S\in\FF}\bW(v^S)$$ 
and hence 
$$v=v' \wedge  (\bigwedge _{S\in\FF}v^S).$$

Note that if we set $v'=[1; 1, \ldots, 1]$ then the construction is essentially the proof that every simple game $v$ is the intersection of weighted games.
 
\item[II)] Let $v_A=[q^A; w_1^A, w_2^A, \ldots, w_n^A]$ and  $v_B=[q^B; w_1^B, w_2^B, \ldots, w_n^B]$ be two weighted games. 
We want to realize $v_A\vee v_B$ as an intersection of $k$ weighted voting systems.

Let $w^I(S)=\sum_{j\in S}w^I_j$ for $I=A$ or $B$ and $\DD=\{S \subset \{1,\ldots,n\}\ :\ w^A(S)<q^A \mbox{ but }w^B(S)\ge q^B\}$. 

If $\DD=\emptyset$ then $v_A\vee v_B=v_A$ and we are done. Now suppose $\DD\ne\emptyset$ and let $u=q^A-\min_{S\in\DD}w^A(S)$
and $\TT=\cap_{S\in\DD}S$.

Suppose that $\TT\ne\emptyset$. Then for each $k\in \TT$, we define 
$$v^k=[q^A; w_1^A, \dots, w_{k-1}^A, w_k^A+u, w_{k+1}^A, \ldots, w_n^A].$$  

Note that if $S$ is winning in $v_A$, then it is winning in $v^k$ for each $k\in \TT$. Now suppose $S$ is winning in $v_B$ but not winning in $v_A$. In this case, $S\in \DD$ by the definition of $\DD$ and hence for each $k\in \TT$, we have $k\in S$ and $w^k(S)=w^A(S)+u\ge q^A$. Hence $S$ is a winning coalition of  $v^k$.

Therefore $\bW(v_A\vee v_B)\subseteq \bW(\bigwedge_{k\in\TT} v^k)$.  If we let $v=v_A\vee v_B$ and $v'=\bigwedge_{k\in\TT} v^k$ and apply the construction I, we have 
$$v_A\vee v_B=(\bigwedge_{k\in\TT} v^k) \wedge (\bigwedge_{S\in\FF}v^S).$$

Note that this method may not yield good results when $\TT$ is not big, as we will see in the next section.
\end{itemize}

We conclude this section by stating the result we just obtained.

\noindent
{\bf Theorem 1}. Let $v_A$ and  $v_B$ be two weighted games with the same set of players. Then 
 $$v_A\vee v_B=(\bigwedge_{k\in\TT} v^k) \wedge (\bigwedge_{S\in\FF}v^S).$$

\section{Upper bound of the dimension of the EU system.}
Recall that $v_{EU}=v_{16}\wedge (v_{65}\vee v_{25})$. Hence the dimension of $v_{EU}$ is $1$ plus the dimension of $v_{65}\vee v_{25}$. In this section, we will obtain a two digit upper bound on the dimension of $v_{65}\vee v_{25}$ based on the populations of the EU members in 2014,2016-18 given in Table 1-4 respectively.  

\subsection{2014 data}
Let $v_A=v_{65}$ and $v_B=v_{25}$. For any $S \subset N$, define $S^c=N\backslash S$. Using the populations given in Table 1, one can check that $\DD=\{\{1,2,3\}^c, \{1,2,4\}^c , \newline \{1,2,5\}^c, \{1,2,6\}^c, \{1,3,4\}^c,\{1,3,5\}^c, \{1,3,6\}^c, \{1,4,5\}^c, \{1,4,6\}^c, \{2,3,4\}^c\}$.
Hence $u=q^A-(w_4+\cdots+w_{28})=33349058$, $\TT=\{7,8,\ldots,28\}$ and $\FF=\{\{1,2,7,8,\ldots,28\}\}$.

Apply Theorem 1 to $v_{65}\vee v_{25}$ so that $v_{EU}=v_{16}\wedge (v_{65}\vee v_{25})$ can be realized as the intersection of the following $24$ weighted games:
$$\begin{bmatrix}
16;& 1,1,1,1,1,1,& 1,& 1,& \ldots,& 1\\
q^A;& w_1, \ldots, w_6,& w_7+u,& w_8,& \ldots,& w_{28}\\
q^A;& w_1, \ldots, w_6,& w_7,& w_8+u,& \ldots,& w_{28}\\
\vdots&  \vdots& \vdots & \vdots & \vdots& \vdots\\
q^A;& w_1, \ldots, w_6,& w_7,& w_8,& \ldots,& w_{28}+u\\
1;& 0,0,1,1,1,1,&0,&0,&\ldots,&0\\
\end{bmatrix}$$

Therefore, the dimension of $v_{EU}$ in $2014$ is at most $24$.

\subsection{A remark}
If we take $v_A=v_{25}$ and $v_B=v_{65}$ using 2014 data, then $\DD$ contains $\{1,2,3,4,5,6\}$, $\{3,4,\ldots,26\}$,  $\{1,2,6,7,\ldots,26\}$,  $\{1,2,3,7,\ldots,27\}$ and hence $\TT=\emptyset$ and our method produces no good upper bound.

\subsection{2016 data}
We take $v_A=v_{65}$ and $v_B=v_{25}$ for the $2016$ data set. Just like the 2014 data set, we have  
$\DD=\{\{1,2,3\}^c, \{1,2,4\}^c,
 \{1,2,5\}^c,$$\{1,2,6\}^c, \{1,3,4\}^c,\{1,3,5\}^c, \{1,3,6\}^c,\newline \{1,4,5\}^c, \{1,4,6\}^c, \{2,3,4\}^c\}$
and $\TT=\{7,8,\ldots,28\}$. However,
$u=35656002$ and $\FF=\{\{1,2,7,8,\ldots,28\},\{1,3,7,8,\ldots,28\}\}$. Hence $v_{EU}$ can be  realized as the intersection of the following $25$ weighted games:
$$\begin{bmatrix}
16;& 1,1,1,1,1,1,& 1,& 1,& \ldots,& 1\\
q^A;& w_1, \ldots, w_6,& w_7+u,& w_8,& \ldots,& w_{28}\\
q^A;& w_1, \ldots, w_6,& w_7,& w_8+u,& \ldots,& w_{28}\\
\vdots&  \vdots& \vdots & \vdots & \vdots& \vdots\\
q^A;& w_1, \ldots, w_6,& w_7,& w_8,& \ldots,& w_{28}+u\\
1;& 0,1,0,1,1,1,&0,&0,&\ldots,&0\\
1;& 0,0,1,1,1,1,&0,&0,&\ldots,&0\\
\end{bmatrix}$$

Therefore, the dimension of $v_{EU}$ in 2016 is at most $25$.

\subsection{2017 data}
For this case, $\DD=\{\{1,2,3\}^c, \{1,2,4\}^c , \{1,2,5\}^c, \{1,2,6\}^c, \{1,3,4\}^c,\{1,3,5\}^c,\newline \{1,3,6\}^c,  \{1,4,5\}^c, \{1,4,6\}^c, \{2,3,4\}^c, \{2,3,5\}^c\}$ and $u=35656002$ 
but we still have $\TT=\{7,8,\ldots,28\}$ and $\FF=\{\{1,2,7,8,\ldots,28\},\{1,3,7,8,\ldots,28\}\}$.

As a result, $v_{EU}$ can be again realized as the intersection of $25$ weighted games as in the case for the $2016$ data set.

\subsection{2018 data}
In this case, everything is identical to that of the $2017$ data set, except  $u=36226115$. Therefore, the dimension of $v_{EU}$ in 2018 is again at most $25$.

\subsection{2018 data without UK}
The United Kingdom has been seriously considering the possibility of leaving the EU during the preparation of this paper.  Therefore, it would be interesting to see what happens to our upper bound if UK is no longer a member of the EU. Using the 2018 data set without UK, we found that $\TT=\{16,17,\ldots,28\}$ and $|\FF|=1348$. Hence, the upper bound of the dimension of $v_{EU}$ then jumps to $1362$.

Also, we would like to point out that Kurz \& Napel lower bound of $v_{EU}$ will also change with the absence of UK. Using the similar method introduced in section 5 of the paper by Kurz \& Napel (2015), we find the following set of losing coalitions with the 'pairwise incompatibility property' (see Kurz \& Napel (2015),  section 4):
\begin{equation*}
    \begin{split}
        \{&\{1, 2, 7, 8, 10, 11, 12, 13, 14, 15, 16, 17, 18, 19, 20, 21, 22, 23, 24, 25, 26, 27, 28\},\\
        &\{2, 4, 5, 9, 10, 11, 12, 13, 14, 15, 16, 17, 18, 19, 20, 21, 22, 23, 24, 25, 26, 27, 28\},\\
        &\{2, 5, 6, 7, 8, 9, 10, 11, 13, 14, 15, 16, 17, 18, 19, 20, 21, 23, 24, 25, 26, 27, 28\},\\
        &\{4, 5, 6, 7, 8, 9, 10, 11, 12, 13, 14, 15, 16, 17, 18, 20, 21, 22, 23, 24, 25, 27, 28\},\\
        &\{1, 5, 6, 7, 9, 10, 11, 12, 13, 15, 16, 17, 18, 19, 20, 21, 22, 23, 25, 26, 27, 28\},\\
        &\{1, 4, 6, 8, 9, 11, 12, 14, 15, 16, 17, 18, 19, 20, 21, 22, 24, 25, 26, 27, 28\},\\
        &\{2, 4, 6, 7, 8, 9, 10, 11, 12, 13, 16, 17, 18, 19, 22, 23, 24, 25, 26, 27, 28\}\}
    \end{split}
\end{equation*}
where the numbers correspond to each country's population ranking in 2018 data. Note that the number $3$ (which represents the UK) is absent. This can be extended by adding the  maximal losing coalition $\{1, 2, 4, 5, \dots, 15\}$ of the $14$ largest countries. Therefore, the lower bound of $v_{EU}$ will increase to $8$ if UK leaves EU.

%

\section{Conclusion}
Kurz \& Napel (2015) used integer linear programing  to estimate the upper bound of the voting system of EU and gets a large bound $13368$.  Our estimation made use of the structure of the voting system to get a much smaller upper bound $25$. Moreover, our estimation indicates that the dimension is sensitive to the populations of the countries.

In general, our method works best if $\TT$ is close to $\{1,\ldots,n\}$. From the fact that $v_1\wedge(v_2 \vee v_3)=(v_1\wedge v_2)\vee(v_1\wedge v_3)$, one may notice that our method can actually be extended to all simple games.

%
%

\end{document}